\documentclass{article}
\usepackage{spconf,amsmath,graphicx}
\usepackage[colorlinks=true, allcolors=blue]{hyperref}
\usepackage{color}
\title{Jeffreys divergence-based regularization of neural network\\output distribution applied to speaker recognition}
\name{Pierre-Michel Bousquet, Mickael Rouvier}
\address{LIA - Avignon University}
\begin{document}

\ninept

\maketitle

\begin{abstract}
A new loss function for speaker recognition with deep neural network is proposed, based on Jeffreys Divergence. Adding this divergence to the cross-entropy loss function allows to maximize the target value of the output distribution while smoothing the non-target values. This objective function provides highly discriminative features. Beyond this effect, we propose  a theoretical justification of its effectiveness and try to understand how this loss function affects the model, in particular the impact on dataset types (i.e. in-domain or out-of-domain w.r.t the training corpus). Our experiments show that Jeffreys loss consistently outperforms the state-of-the-art for speaker recognition, especially on out-of-domain data, and helps limit false alarms.
\end{abstract}
% \noindent\textbf{Index Terms}: Speaker recognition, deep learning, loss function

%%%%%%%%%%%%%%%%%%%%%%%%%%%%%%%%%%%%%%%%%%%%%%%%%%%%%%%%%%%%%%
\section{Introduction}
\label{sec:Introduction}
In recent years, Deep Neural Networks (DNN) have achieved remarkable performance in speaker recognition (SR) compared to the traditional i-vector/PLDA framework~\cite{Dehak11_a}. Proposing an original and discriminant voice representation, the DNN can be seen as a complex function that maps the audio to a vector (i.e. speaker embedding). The loss function plays an important role to determine the DNN parameters during the learning phase. Hence, choosing the suitable loss function is crucial to estimate discriminant parameters, achieve a good accuracy and avoid drawbacks that typically occur with deep learning (e.g. overfitting, miscalibration~\cite{Guo2017},...).

Two major dimensions of research around loss functions may be found in the machine learning literature. Some of them are based on classification (softmax cross-entropy loss, center loss~\cite{Wen2016}), while others achieve representation learning (contrastive loss~\cite{Heigold2014}, triplet loss~\cite{Chao2017,Bredin2017}, circle loss~\cite{Sun2020C}, barlow twins~\cite{Zbontar2021,mohammadamini2022barlow}).
% simsiam~\cite{}...).
%, sphereface~\cite{liu2018sphereface}, cosface~\cite{},  arcface~\cite{}
However, both types of loss functions suffer from major issues: the triplet loss for representation learning, for instance, exhibits a combinatorial explosion in the number of possible triplets, especially for large-scale datasets, leading to a drastically increased number of training steps. On the other hand, loss functions based on classification may see a linear increase of the size of the linear transformation matrix with the number of identities; the learned features are separable for the closed-set classification problem but not discriminative enough for the open-set SR problem.

The softmax cross-entropy loss is typically good for optimizing the inter-class difference (i.e., separating different classes) but not for reducing the intra-class variation (i.e., making classes more compact). To address this issue, many loss functions have been proposed, attempting to minimize the intra-class variation: sphereFace~\cite{liu2018sphereface}, cosFace~\cite{Wang18,Wang2018CosFaceLM},  arcFace~\cite{Deng2019arcface}.
Based on angular distances, which tend to be the state of the art in SR, they can be seen as \textit{embedding losses} as they all rely on the generic softmax loss function.
Therefore, completing the cross-entropy loss by a regularizer of the output distribution could improve all these configurations.

Two goals can be identified for the objective function of a DNN: reflect the true objective of the model learning (to discriminate the training speakers) and the real objective of the system (to avoid overfitting, in order to generalize well to new data).
These two goals are independent, even contradictory, and the regularizer will have to find a trade-off between them.

In this paper, a new regularizer of output distributions, the Jeffreys loss, is presented in Section~\ref{sec:Jeffreys}. Before that, Section~\ref{sec:Related_Work} and~\ref{sec:Probing} present the different loss functions, probe the output distribution and justify our approach.
Results of wide and deep ResNet systems on speaker verification tasks are analyzed in Section~\ref{sec:Experiments} and conclusions are provided in Section~\ref{sec:Conclusion}.

%%%%%%%%%%%%%%%%%%%%%%%%%%%%%%%%%%%%%%%%%%%%%%%
\section{Loss functions for speaker recognition}
\label{sec:Related_Work}

Regularizing the output distribution of deep and wide neural
networks has long been unexplored. Since, many
studies have shown the benefits of loss function regularizers.
First recall that, for each training example, the model computes a conditional distribution over labels $k\in\{1...K\}$ given the x-vector $x$ through a softmax function: $p\left(  k|x\right)  =\exp\left(  z_{k}\right)  /{\textstyle\sum_{i=1}^{K} }\exp\left(  z_{i}\right)  $ where $z_{i}$ are the logits. In SR, logits are actually dot product~\cite{Snyder2018} or cosine (coupled with the angular softmax loss function~\cite{Wang18,Deng2019arcface}), eventually shifted and scaled (penalty margin~\cite{Wang18}, temperature scaling). The most currently used loss in SR is the cross-entropy in the case of hard target (a single ground-truth label equal to $1$ for $k$, otherwise $0$) and minimizing the cross entropy is equivalent to maximizing the logit of the correct label. Omitting the dependence of $p$ on example $x$ and denoting $p\left(k|x\right)  $ by $p_{k}$, the cross-entropy loss for the example below is equal to $\mathcal{L}_{CE}=-\log\left(  p_{k}\right)$.

In SR, learning enhancement by temperature scaling of the logit is typically set to low values, fastening convergence and limiting overfitting. It has been also shown that it provides better calibrated scores~\cite{Guo2017}. A popular countermeasure against overfitting is addition of a regularization term to the objective function. Label-smoothing~\cite{Szegedy2015LabelSmoothing}, widespread in many fields (image classification, language modeling, machine translation, speech recognition, digit recognition, ...), replaces a "hard" target label objective by a "softened" one, playing on non-target values of the output distribution. This leads to add to $\mathcal{L}_{CE}$ a weighted term $\mathcal{L}_{LS}=\tfrac{1}{K-1}\sum_{i\neq k}\log p_{i}$. Some variants have been proposed~\cite{Pereyra2017regularizing,Lienen_Hüllermeier_2021}. Smoothing the labels by other ways, such as virtual adversarial training~\cite{Miyato2016}, adding label noise~\cite{Reed2014,Xie2016} or addressing class imbalance~\cite{Lin2017}, has also been effective in preventing overfitting and, thus, improving generalization. Each time, cooperation of these methods with weight decay~\cite{Krizhevsky2009} must be analyzed and overcome.

In the following, the benefits of the output regularizing and its ability to better achieve the two goals defined in the introduction are analyzed and justified, more thoroughly than in the literature, leading us to propose a more comprehensive regularizer of the output distribution for SR.
%%%%%%%%%%%%%%%%%%%%%%%%%%%%%%%%%%%%%%%%%%%%%%%
\section{Probing the output distributions}
\label{sec:Probing}
\subsection{Better discriminate the training speakers}
\begin{figure}[t]
\centering
\includegraphics[scale=0.45]{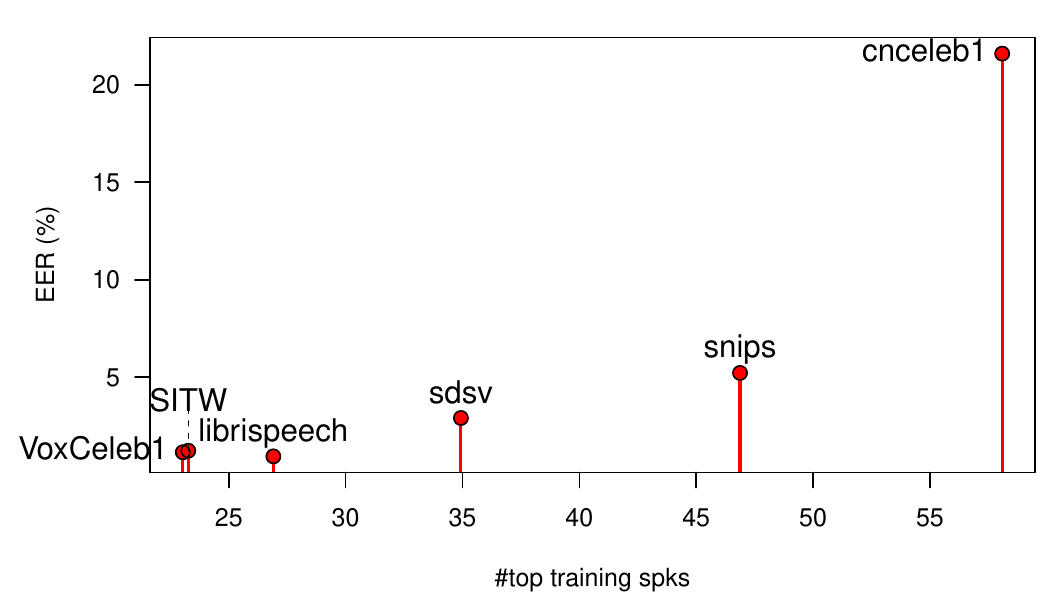}
\vspace{-0.25cm}
\caption{Equal error rates on the six evaluations presented in~\cite{Bousquet2022} as a function of the average number of top training
speakers ($x$-values). See Section~\ref{sec:Probing} and~\cite{Bousquet2022}
for more details.}%
\label{fig_nTop_EER.pdf}%
\end{figure}
\begin{figure}[t]
\centering
\includegraphics[scale=0.5]{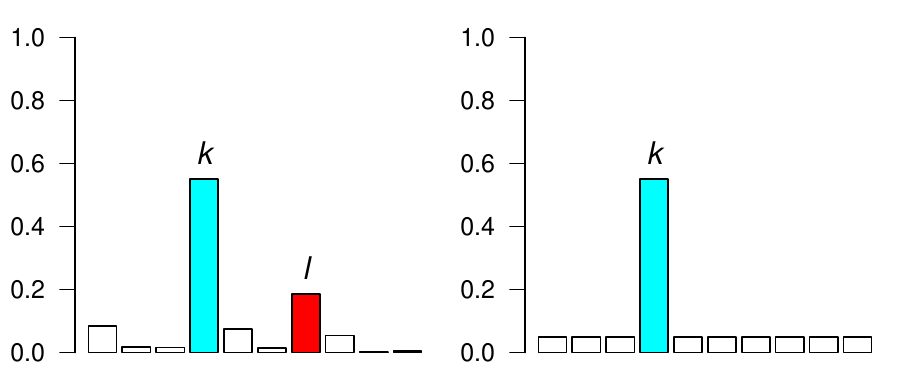}
\vspace{-0.25cm}
\caption{Two examples of output softmax distribution for an example of the
$k^{\text{th}}$ training speaker. As their $k^{\text{th}}$ values are equal,
both yield the same value of one-hot-target cross-entropy loss.}%
\label{fig_two_outputs}%
\end{figure}

In~\cite{Bousquet2022} the notion of \textit{top training speakers} is introduced, that is, given an utterance of a speaker
unknown to the system, the dominant labels of its output distribution. To
summarize the approach, once the x-vector of a test utterance is extracted,
the softmax of the last layer is computed and the labels with the highest values are retained. The corresponding training speakers are those who are the most involved in the modeling of this utterance.

To take this study a step further, Figure~\ref{fig_nTop_EER.pdf} reports, for
each evaluation presented in~\cite{Bousquet2022},
the average number of top training speakers ($x$-values) of its utterance set
and the equal error rates (EER) computed on its trial set ($y$-values).
From Figure~\ref{fig_nTop_EER.pdf}, it is clear that an increase in the number of top training speakers correlates with an increase in the observed EER.
The network, trained as a classifier for the training set, proceeds with new
speakers by similarity to it. Too many top-speakers reveal some difficulties
in the system to model the data and, therefore, the SR system performs better when it succeeds in modeling test data with a low number of main training speakers.
This shows that performance of a system on a domain (and similarity of the latter with the training data) can be predicted by only probing the last layer of the data, but it also highlights that the ability of a system to generalize is measurable by the entropy between test and training output distributions.

Figure~\ref{fig_two_outputs} illustrates 
the link with the entropy between the training data outputs.
The figure shows two examples of output softmax distribution
for an example of the $k^{\text{th}}$ training speaker (the training
speaker sample size is limited for readability). As expected, the target
value $p_{k}$ is maximal for both cases, and equal, so that the cross-entropy
loss induced by both examples is identical.

However, on the left, a non-negligible "foreign" (non-target) value for the $l^{\text{th}}$ label (red bar)
reduces the entropy between the output distributions of the $k^{\text{th}}$ and $l^{\text{th}}$ training speakers
(which can be measured by the symmetric Kullback-Leibler divergence).
This concern is not taken into account by the cross-entropy loss.
The distribution on the right of the Figure, where the foreign values
$\left\{  p_{i}\right\}  _{i\neq k}$ follow a uniform distribution, increases the entropy between the output distributions of training speakers $k$ and $l$, thus helping to better discriminate them
and to achieve the first goal outlined in the previous section. Techniques
such as label-smoothing rely on this observation and attempt to equalize the non-target
values of the output.

\subsection{Avoid overfitting}
But label-smoothing could lead to overfitting, by taking too much into account the specificity of the training data.
% by increasing the overall
% pairwise-entropy between speakers samples and, thus, optimizing the model to the
% specificity of the training data.
This finding is contrary to what is usually claimed and empirically justified~\cite{Muller2019,Pereyra2017regularizing}.

In what
follows, we propose to better explain why smoothing the non-target labels
also respects the second goal of the learning phase: 
avoid overfitting. 
Let $q_{1}$ and $q_{2}$ denote the two output distributions displayed in Figure~\ref{fig_two_outputs}. Now consider
an SR domain and its output set $\mathcal{P}$. The mean entropy between
$\mathcal{P}$ and $q_{1}$ or $q_{2}$ can be estimated by the expectations
%$\left\{  \mathbf{E}_{p\in\mathcal{P}}\left[  D_{KL}\left(  p||q_{i}\right)
%\right]  \right\}  _{i=1,2}$.
$\mathbf{E}_{p\in\mathcal{P}}\left[  D_{KL}\left(  p||q_{i}\right)  \right]
$, $i=1,2$. This amounts to comparing $\mathbf{E}_{p\in\mathcal{P}%
}\left[  p\right]  .\log q_{i}$ where '$.$' denotes the dot product. When
$\mathcal{P}$ is far from the training domain, the values of $p$ tend to be spread across many labels (as observed above and in Figure~\ref{fig_nTop_EER.pdf}), so that $\mathbf{E}_{p\in\mathcal{P}}\left[  p\right]
$ tends towards the uniform distribution, many top speakers inducing smoother
output.\ Therefore, $\mathbf{E}_{p\in\mathcal{P}}\left[  D_{KL}\left(
p||q_{2}\right)  \right]  $ should probably be lower than
$\mathbf{E}_{p\in\mathcal{P}}\left[  D_{KL}\left(  p||q_{1}\right)  \right]  $.

In other words, by smoothing the non-target values of the training set
output distributions, redundancy between training speakers is reduced but,
also, the entropy of out-of-domain data a posteriori of the model.
% From a geometrical point of view, the representation space is expanded
% to better host and model heterogeneous data,
% according to the available training material.
% This contributes to limit overfitting, thus to improve generalization.

%%%%%%%%%%%%%%%%%%%%%%%%%%%%%%%%%%%%%%%%%%%%%%%
\section{Jeffreys-based loss function}
\label{sec:Jeffreys}

\begin{figure*}[t]
\centering
\includegraphics[scale=0.5]{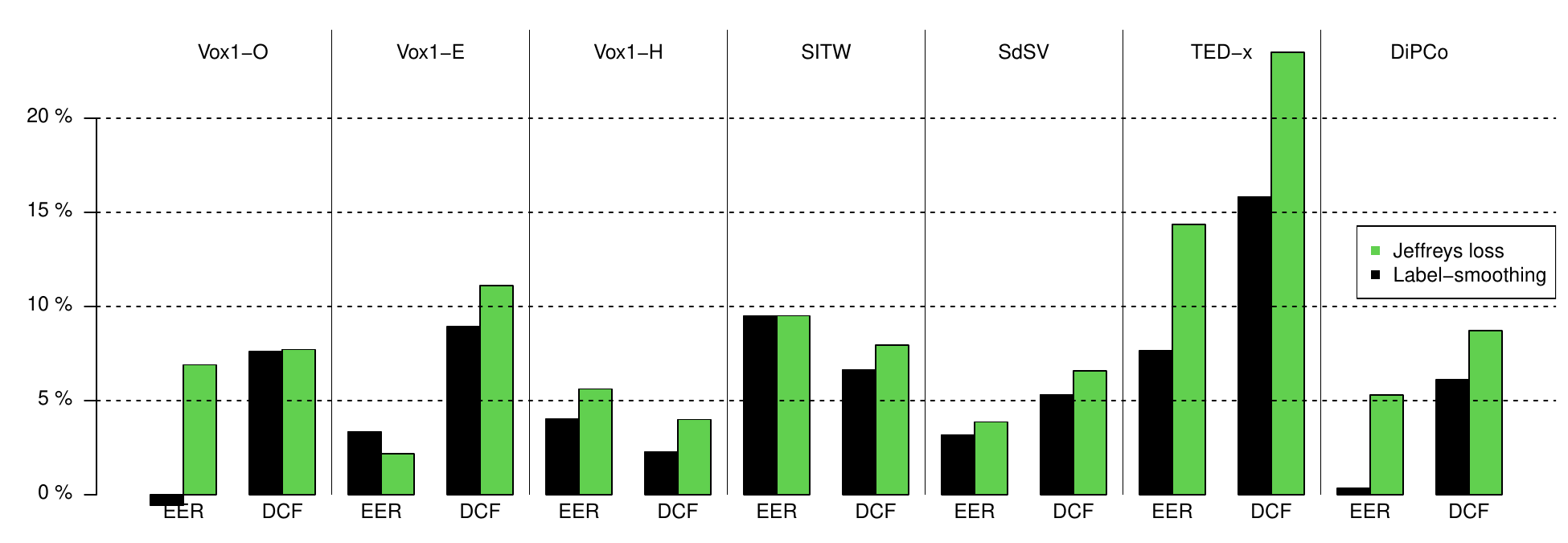} 
\vspace{-0.25cm}
\caption{Relative gains between the 'hard'-target cross entropy loss (row 1 of
Table~\ref{tbl:results}) and two regularizers: label-smoothing (black bars and row 3 of Table~\ref{tbl:results}) then
Jeffreys loss (green bars, row 4 of the Table) in terms of EER and DCF.
The VoxCeleb1 evaluations are fully in-domain, SITW is relatively in-domain, the others are out-of-domain.}
\label{fig_results}%
\end{figure*}

\begin{table*}[ht]
\caption{Comparison of  the cross entropy loss with two regularizers: label-smoothing (with or without weight decay) and the new Jeffreys-based loss function.}%
\label{tbl:results}
\center
%\resizebox{\columnwidth}{!}{
\resizebox{7in}{!}{
\begin{tabular}[|c]{|l|c|c|c|c|c|c|c|c|c|c|c|c|c|c|}
\hline
\textbf{System}
& \multicolumn{2}{c|}{\textbf{VoxCeleb1}}
& \multicolumn{2}{c|}{\textbf{VoxCeleb1}}
& \multicolumn{2}{c|}{\textbf{VoxCeleb1}}
& \multicolumn{2}{c|}{\textbf{SITW}}
& \multicolumn{2}{c|}{\textbf{SdSV}}
& \multicolumn{2}{c|}{\textbf{TED-x Spanish}}
& \multicolumn{2}{c|}{\textbf{DiPCo}}\\

& \multicolumn{2}{c|}{\textit{-O Cleaned}}
& \multicolumn{2}{c|}{\textit{-E Cleaned}}
& \multicolumn{2}{c|}{\textit{-H Cleaned}}
& \multicolumn{2}{c|}{\textit{core-core}}
& \multicolumn{2}{c|}{\textit{task 2}}
& \multicolumn{2}{c|}{}
& \multicolumn{2}{c|}{}\\

 & EER & DCF & EER & DCF & EER & DCF & EER & DCF & EER & DCF & EER & DCF & EER & DCF\\\hline
AAM-Softmax &0.93&0.095&1.01&0.114&1.76&0.170&1.15&0.101&3.46&0.295&2.33&0.178&5.85&0.368\\
AAM-Softmax + LS with weigth-decay &0.94&0.087&1.05&0.112&1.80&0.170&1.15&0.101&3.63&0.296&2.26&0.140&5.60&0.356\\
AAM-Softmax + LS w/o weigth-decay &0.93&0.087&0.98&0.104&1.69&0.166&1.04&0.094&3.36&0.280&2.15&0.149&5.83&0.346\\
AAM-Jeffreys  &0.86&0.087&0.99&0.102&1.66&0.164&1.04&0.093&3.33&0.276&1.99&0.136&5.54&0.336\\
\hline
\end{tabular}
}\end{table*}

As shown above, regularizing the network w.r.t. the two goals defined
in Section~\ref{sec:Introduction} can be done by moving the non-target output
distribution $p$ closer to the uniform distribution $u$.
The most complete entropy measure between $p$ and $u$ is the symmetric
Kullback-Leibler divergence (also referred to as Jeffreys divergence). This
divergence takes into account the entropy of $u$ a posteriori of $p$ (as done
in label-smoothing) but, above all, the one of $p$ a posteriori of $u$.

To apply Jeffreys divergence on the non-target output softmax values $\left[
p_{i}\right]  _{i\neq k}$, this sub-vector is $L1$-normalized to become a
distribution (thus divided by $1-p_k$), then the Jeffreys divergence-based loss between it and the
uniform distribution $u$, equal to $\frac{1}{K-1}$ for all its $K-1$ values, is
computed:

\begin{equation}
\mathcal{L}_{J}=D_{KL}\left(  u|\left[ \frac{p_{i}}{1-p_{k}}\right]
_{i\neq k}\right)+D_{KL}\left(
\left[\frac{p_{i}}{1-p_{k}}\right]  _{i\neq k}|u\right)
\label{eq_Jeffreys_1}%
\end{equation}
After simplification, this is equal to :%

\begin{equation}
\mathcal{L}_{J} = -\frac{\sum_{i\neq k}\log p_{i}}{K-1} + \frac{\sum_{i\neq k}p_{i}\log p_{i}}{1-p_{k}}
\label{eq_Jeffreys_2}%
\end{equation}

The final loss is a weighted sum of the cross-entropy and Jeffreys losses :%
\begin{equation}
\mathcal{L}=\mathcal{L}_{CE}+\alpha\mathcal{L}_{J}
\label{eq_final_1}
\end{equation}
where $\alpha$ is a scalar.
The second term of Equation
$\ref{eq_Jeffreys_2}$ is a little bit 'hard' as it inserts $p_{k}$ instead of $\log\left(  p_{k}\right)$ inside the loss function. To alleviate this effect, the two terms of Jeffreys loss are independently
weighted~\footnote{To facilitate further research, the code of the loss-function is available on \url{https://github.com/mrouvier/jeffreys_loss}}:

\begin{equation}
\mathcal{L}=-\log\left(  p_{k}\right)  -\alpha\frac{\sum_{i\neq k}\log p_{i}}{K-1} +\beta\frac{\sum_{i\neq k}p_{i}\log p_{i}}{1-p_{k}}
\label{eq_final_3}%
\end{equation}

This loss can be rewritten by
using the label-smoothing loss:%
\begin{equation}
\mathcal{L}=\mathcal{L}_{CE}+\alpha\mathcal{L}_{LS}
+\beta\frac{\sum_{i\neq k}p_{i}\log p_{i}}{1-p_{k}}
\label{eq_final_2}%
\end{equation}

This result shows that label-smoothing is only a part of the divergence
between non-target values and uniform distribution. The last term of Equation
$\ref{eq_final_2}$ simultaneously forces the non-target values to be as
uniform as possible and completes the cross-entropy loss objective (thanks to
the denominator $1-p_{k}$)~\footnote{Let us note that
\cite{Pereyra2017regularizing} includes a term $p_{i}\log p_{i}$ in a loss function, but without the denominator, including the target value and which does not improve performance when combined with label-smoothing.}.

%%%%%%%%%%%%%%%%%%%%%%%%%%%%%%%%%%%%%%%%%%%%%%%
\section{Experiments}
\label{sec:Experiments}
%%%%%%%%%%%%%%%%%%%%%%%%%%%%%%%%%%%%%%%%%%%%%%%
\subsection{Experimental setup}
\label{subsec:Experimental_setup} 
The $x$-vector extractor used in this paper is a variant based on ResNet-34. The extractor was trained on the development part of the Voxceleb 2
dataset~\cite{VoxCeleb2_chung2018}, cut into 4-second chunks and augmented with noise, as described in~\cite{Snyder2018} and available as a part of the Kaldi-recipe. It contains about 1M segments (+ 4M augmented) of 5994 speakers. As input, we used 60-dimensional filter-banks. The speaker embeddings are 256-dimensional and the loss is the angular additive margin with temperature scaling equal to 30 and margin equal to 0.2. The sizes of the feature maps are 128, 128, 256 and 256 for the four ResNet blocks. We use stochastic gradient descent with momentum equal to 0.9, a weight decay equal to 2.10$^{-4}$ and initial learning rate equal to 0.2. The implementation is based on PyTorch.
For scoring, the $x$-vectors are centered by subtracting the overall mean of the training dataset, then the cosine metric is applied. Despite the shift between some tests and training data, no domain adaptation technique is performed in order to fairly compare the effects of regularizers.

The relevance of the methods is tested on seven datasets: VoxCeleb1-O, E and H (cleaned versions)~\cite{VoxCeleb_Nagrani_2017,voxCeleb_Xie_2019},
Speakers In The Wild (SITW) core-core task~\cite{SITW_mclaren2016speakers},
the Short duration Speaker Verification (SdSV) challenge Task 2 (a text-independent SR evaluation based on the DeepMine dataset~\cite{deepmine2018odyssey,deepmine2019asru}, comprised of Persian-native and some English-non native utterances),
DiPCo~\cite{rouvier2022_lrec} (a far-field speaker verification corpus issued from DiPCo corpus)
and TED-x Spanish.
% ~\footnote{\scriptsize \url{http://dipco-sre.univ-avignon.fr/downloads/tedx_spanish.tar.gz}}.
% TED-x Spanish dataset is derived from the public-available TED-x Spanish. TED-x Spanish is a dataset created from TED talks in Spanish and it aims to be used in the Automatic Speech Recognition Task. TED-x Spanish contains spontaneous speech of several expositors in TED-x events (142 speakers). The dataset is a gender unbalanced corpus (102 males and 40 females) of 24 hours of duration. Transcription and segmentation files for the corpus are provided. For the derived corpus, we extracted enrollment and test segment from the segmentation. The segments have a duration range between 3 and 10 seconds. We randomly selected 2 millions of pairs (1.600.000 non-targets pairs and 400.000 targets pairs). In order to create a more challenging dataset, we only selected pairs with the same gender. The derived corpus is freely available here~\footnote{XXXX}.
The latter is derived from the public-available TED-x Spanish dataset, created from TED talks in Spanish and aiming to be used in the Automatic Speech Recognition Task.
For the derived corpus, the segments have a duration range between 3 and 10 seconds and we randomly selected 2M pairs (1.6M non-targets pairs and 0.4M targets pairs), all with the same gender.
% The derived corpus is freely available here~\footnote{XXXX}.

The last three evaluations are "out-of-domain", due to mismatch of language or recording conditions.
% It may be noticed that our 5994-size corpus of  VoxCeleb2 training speakers includes a non-negligible share of Indian speakers (427), which could slightly limit the domain shift with SdSV Task 2.

The results are reported in terms of Equal Error-Rate (EER) and normalized minimal detection cost (DCF) with the probability of a target trial set to 0.01 and the cost of miss-detection and false alarm set to 1.

%%%%%%%%%%%%%%%%%%%%%%%%%%%%%%%%%%%%%%%%%%%%%%%
\subsection{Results}
\label{subsec:Results}
\begin{figure}[t]
\centering
\includegraphics[scale=0.5]{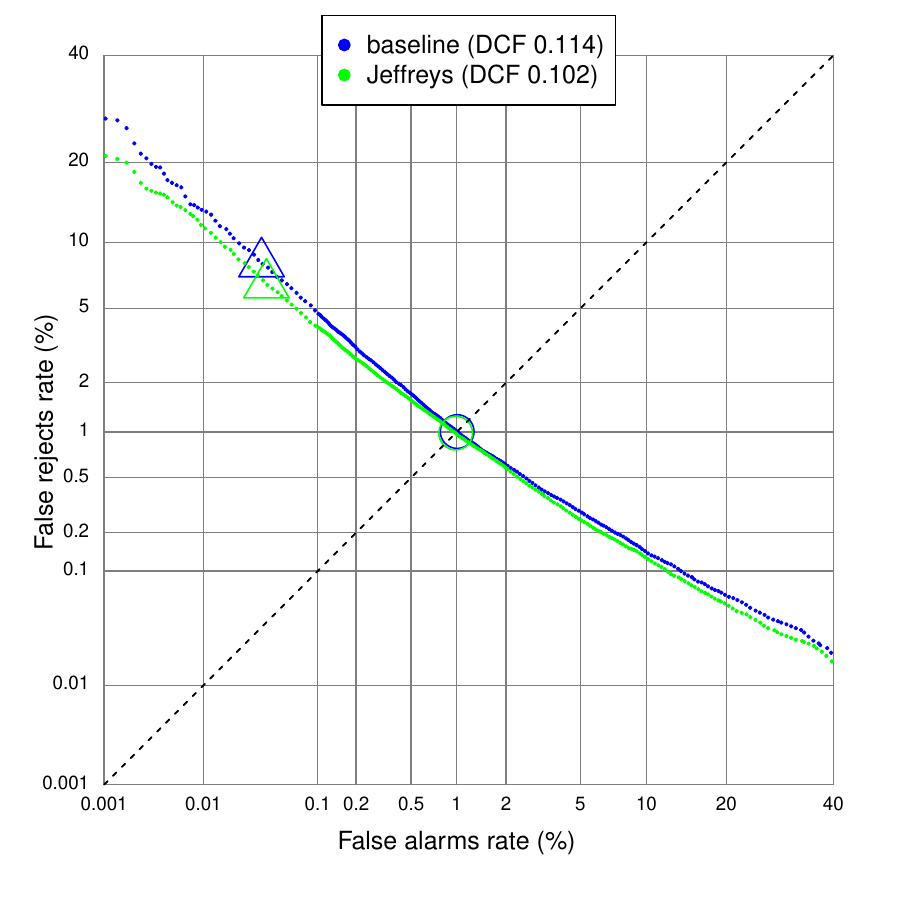}
\vspace{-0.5cm}
\caption{DET curves for VoxCeleb1-E. The circles are the EERs, the triangles the DCFs.}
\label{fig_DET_curve_Vox1E}%
\end{figure}
\begin{figure}[t]
\centering
\includegraphics[scale=0.5]{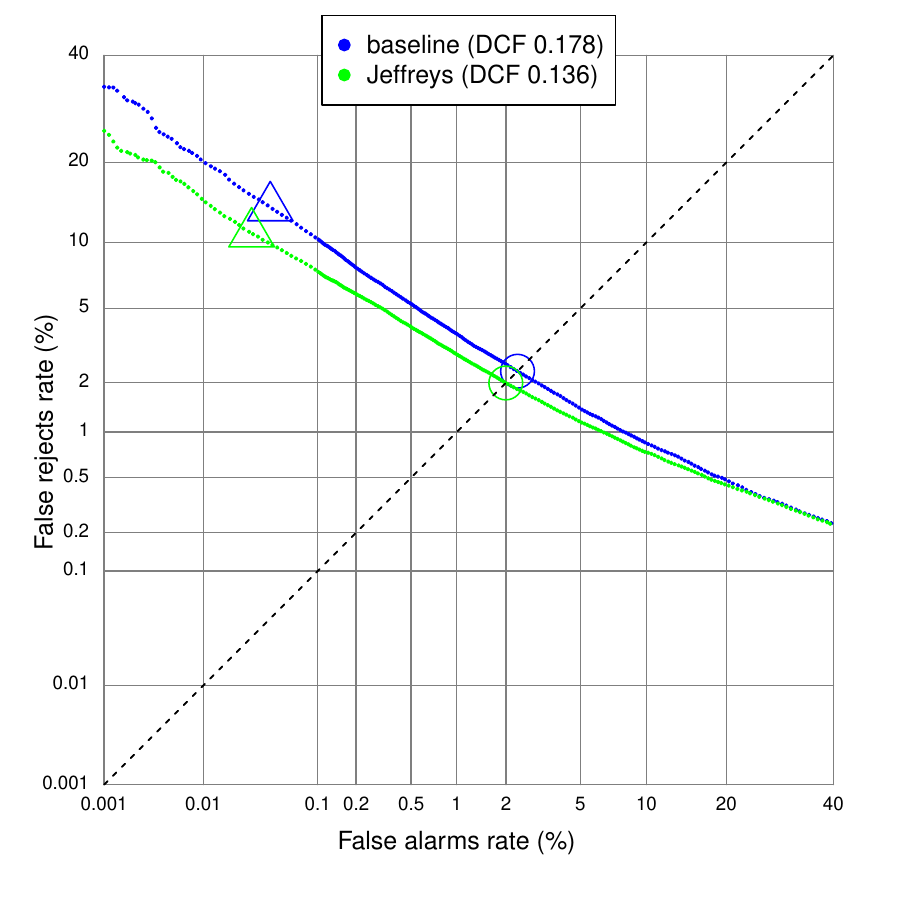}
 \vspace{-0.5cm}
\caption{DET curves for TED-x Spanish.}
\label{fig_DET_curve_TED-x}%
\end{figure}

Table~\ref{tbl:results} shows the evaluation results.
Row 1 reports results of the system learned
with the cross-entropy loss function.
This system is stated here as the baseline.
Row 2 reports results with label-smoothing. As weight decay is
also a regularization technique and could interfere and clash with label-smoothing, row 3 reports results
obtained without weight decay. Comparing the first three rows show that label-smoothing improves performance provided that weight decay is disabled.
Let us note that the best weight for label-smoothing, in terms of performance, was equal to $\alpha=0.1$.
Row 4 reports results with the proposed Jeffreys-based loss
function. No weight decay is applied. The best weights of Eq.~\ref{eq_final_3} were $\alpha=0.1$ and $\beta=0.025$.

To more easily assess the benefits of the regularizers, Figure~\ref{fig_results} visualizes the relative gain between the hard-target cross entropy and the two approaches: label-smoothing (black bars), then Jeffreys loss (green bars).
On VoxCeleb1 evaluations, which are fully in-domain, the gains of performance confirm the ability of the non-target label smoothings to better fulfill the first goal ("\textit{to discriminate the training speakers}").
These gains are comparable to those observed in other
fields~\cite{Szegedy2015LabelSmoothing,Pereyra2017regularizing} and sometimes even greater.
The gain on SITW is significant with both methods.
On out-of-domain evaluations (SdSV, TED-x Spanish, DiPCo), the significant gains of performance demonstrate that the regularizers also achieve the second goal ("\textit{to generalize well}").

The new Jeffreys-based loss function always yields better accuracy than label-smoothing (except for VoxCeleb 1-E EER), especially for out-of-domain evaluations.
The proposed approach tackles overfitting and can be considered more robust to domain mismatch than label-smoothing.
In particular, the Jeffreys regularizer provides significant gains in terms of DCF, even spectacular for out-of-domain evaluations. The method helps to regularize the upper tail of the non target score distribution,
as illustrated by detection error tradeoff (DET) plots in
Figures~\ref{fig_DET_curve_Vox1E} and~\ref{fig_DET_curve_TED-x}.
This ability to produce more calibrated scores is of the utmost importance for critical applications (forensic, security) where false alarms must be severely penalized.

%%%%%%%%%%%%%%%%%%%%%%%%%%%%%%%%%%%%%%%%%%%%%%%
% \vspace{-0.5cm}
\section{Conclusion}
\label{sec:Conclusion}
In this work, several topics about regularization of SR neural network outputs are addressed. First, label-smoothing deserved to be tested for SR. Its results are reported, tested on various evaluations more or less far from the training domain. Second, this regularization is claimed to avoid overfitting, but this outcome is only checked empirically. Here, we show why this assertion is paradoxical a priori, and propose a more theoretical justification showing how such soft target approaches can simultaneously achieve both objectives defined above: discriminating the training speakers and generalizing well to new data. These investigations lead us to propose a new loss function for SR DNN, more comprehensive and robust than label-smoothing, which improves accuracy of the recognition, in particular for the cases of domain mismatch and critical applications. Moreoever, this new loss function is compatible with all recent techniques used in SR: sphereFace, cosFace, arcFace...

These results seem promising enough to propose future work testing this new loss function on other fields (image classification, machine translation, language modeling, speech recognition) on which label-smoothing has proven to be effective.

\section{Acknowledgements}
This work was supported by the VoicePersonae project ANR-18-JSTS-0001, granted access to the HPC resources of IDRIS under the allocation 2022-AD011013257R1 made by GENCI and supported by the ANR agency (Agence Nationale de la Recherche).
Many thanks to Teva Merlin for proofreading this work.
% This work was supported by the VoicePersonae project ANR-18-JSTS-0001.
% This work was granted access to the HPC resources of IDRIS under the allocation 2022-AD011013257R1 made by GENCI and was supported by the ANR agency (Agence Nationale de la Recherche).
% We thank Teva Merlin for his helpful re-reading on this work.

\bibliographystyle{IEEEtran}
\bibliography{biblio_pmb.bib}
%\bibliography{biblio_pmb.bib}

\end{document}